%% file: path_nh1_1_.tex
 \documentclass[graybox]{svmult}

\usepackage{mathptmx}       
\usepackage{helvet}         
\usepackage{courier}        
\usepackage{type1cm}        
\usepackage{makeidx}         
\usepackage{graphicx}        
\usepackage{multicol}        
\usepackage[bottom]{footmisc}

\makeindex

\begin{document}

\title*{ Green's function of a  general PT-symmetric non-Hermitian  non-central potential
 }


\author{Brijesh Kumar Mourya and Bhabani Prasad Mandal }

\institute{Brijesh Kumar Mourya \at
Department of Physics, Banaras Hindu University, Varanasi-221005,India.\email{brijeshkumarbhu@gmail.com}
\and Bhabani Prasad Mandal \at Department of Physics, Banaras Hindu University, Varanasi-221005,India.
\email{bhabani.mandal@gmail.com } }


\maketitle


\abstract{
We study the path integral solution of a system of particle moving in
certain class of PT symmetric non-Hermitian and  non-central potential. The
 Hamiltonian of the system is converted
to a separable Hamiltonian of Liouville type in parabolic coordinates and
is further mapped into a Hamiltonian corresponding to two 2- dimensional
simple harmonic oscillators (SHOs). Thus the explicit Green's functions for a general non-central PT symmetric
non hermitian potential are calculated in terms of that of 2d SHOs.
The entire  spectrum for this three dimensional system is shown to be always real leading to the fact that the system remains in unbroken PT phase all the time}.



\section{ Introduction}
\label{s1}


Feynman's  path integral (PI) approach to quantum mechanical systems is an elegant 
formalism and powerful in semi-classical calculations \cite{fey}. PI formalism which is generally tied to the 
Lagrangian formalism of mechanics is an extremely powerful technique in quantum mechanics. In a class of
problems it provides the Green's function with tremendous ease and also provides valuable insight into the relation
between classical and quantum mechanics. Green's functions which in mathematics are to solve 
non-homogeneous boundary value problems are the backbone of any calculations of physical quantities in quantum
field theory \cite{adas}. Thus it is extremely important for any physical theory that the Green's functions are well defined. The
purpose of the present work  is to discuss the PI formulation of a general non-central, combined parity (P) and time
reversal (T) symmetric non-Hermitian system  in 3d by calculating the explicit Green's functions for such a system.
Consistent quantum theory with real energy eigenvalues, unitary
time evolution and
probabilistic interpretation for PT symmetric 
non-Hermitian theories in a different Hilbert space equipped with positive
 definite inner
product has been the subject of intrinsic research in frontier physics over the last one
and half decades \cite{pt}. Such non-Hermitian PT symmetric systems generally exhibit PT phase transition 
or more specifically a PT breaking transition \cite{ptp} which has been realised experimentally \cite{pto}.
In-spite of huge success and wide applicability \cite{pta,pta0,pta1,pta2,pta3,pta4} of this field the study of non-Hermitian quantum mechanics
is mostly restricted to one dimensional or central potentials in higher dimension potential problems.
In this article we consider a  general and very important non-central  PT symmetric non-Hermitian system
in the PI formulation. First we show that how the Hamiltonian corresponding to this potential is reduced to a    separable Hamiltonian of Liouville type \cite{fu} in a different coordinate system. This  further enables us to
map the system into  two non-interacting 2d harmonic oscillators with the appropriate choice of coordinates. We then calculate the Green's functions of the system in terms of the Green's functions of 2d harmonic oscillators.
Further we write the Hamiltonian in terms of appropriate creation and annihilation operators to calculate the
energy eigenvalues of this non-central non-Hermitian system. We find that the energy eigenvalues are always real as long as the parameters in the potential are real. This indicates that system is always in unbroken PT phase.

Now we present the plan of the paper. In Sec. \ref{s2} we calculate the Green's function for the system in terms of
SHO Green's functions. The reality of the spectrum is shown in Sec. \ref{s3}.  Sec. \ref{s4} is kept for concluding remarks.

\section{Green's functions for the PT-symmetric non-central potential }
\label{s2}
We consider a system described by a general non-central non-Hermitian potential in 3-dimension in spherical polar coordinates  as
\begin{equation}
H= \frac{P_r^2}{2m}+\frac{P_\theta^2}{2mr^2} +\frac{P_\phi^2}{2mr^2 sin^2\theta}+V(r,\theta)
\end{equation}
where the non-Hermitian potential is
\begin{equation}
V(r, \theta ) = -\frac{ \alpha}{r} + \frac{ B\hbar^2}{2mr^2\sin^2 \theta }
+ \frac{ iC\hbar^2\cos\theta }{2mr^2\sin^2 \theta}.
\label{pot}
\end{equation}
 $\alpha,\ B $ and $C$ are real constants. 
It is straight forward to check that this non-Hermitian system is PT symmetric, where in 3-d in spherical polar coordinates the parity transformation is defined as, $ r\rightarrow r ;\ \ \ \theta \rightarrow \pi -\theta ,\ \ \ \phi\rightarrow \phi+2\pi $.This particular potential is very important as the Coulomb and the ring- shaped potentials are particular cases  of 
this potential. For $C =0$ this potential becomes Hartman's ring shaped potential which was originally proposed to 
model Benzene molecule \cite{hat}.
To proceed with this Hamiltonian we first consider the
 most general Hamiltonian of Liouville type written as 
\begin{equation}
H= \frac{1}{ V_1(q_1) +V_2(q_2)}\left\{ \frac{1}{2mW_1(q_1)} p_1^2 + \frac{1}{
2mW_2(q_2)}p^2_2 + U_1(q_1) +U_2(q_2) \right\} 
\label{21}
\end{equation}
This is reduced to a simpler form by using some canonical
transformation 
and redefining $U$ 's and $V$'s
 as \cite{fu}
\begin{equation}
H= \frac{1}{ V_1(q_1) +V_2(q_2)}\left\{ \frac{1}{2m} p_1^2 + \frac{1}{
2m}p^2_2 + U_1(q_1) +U_2(q_2) \right\}
\label{22}
\end{equation}
The time independent Schrodinger equation corresponding to this Hamiltonian is then written as
$$  \hat{H}_T\psi =0 $$ with a total Hamiltonian is defined as
\begin{equation}
\hat{H}_T=  \frac{1}{2m} p_1^2 + \frac{1}{
2m}p^2_2 + U_1(q_1) +U_2(q_2)- E(V_1(q_1) +V_2(q_2))
\label{22}
\end{equation}

Now, to obtain the path integral for this Hamiltonian $\hat{H}_T$, let us  consider
the evaluation of this operator for a arbitrary parameter $\tau$ :
\begin{eqnarray}
\left <q_{1b},\, q_{2b} \left |\exp \left \{-i  \frac{ \hat{H}_T\tau}{\hbar}\right \}
\right | q_{1a}, \, q_{2a} \right >= 
\left < q_{1b} \left | \exp \left \{ -( \frac{ i}{\hbar})\left [ \frac{ 1}{2m}\hat{p}_1^2 +U_1(q_1) -EV_1(q_1)
\right ] \tau \right \} \right | q_{1a} \right > \nonumber \\
\times   \left < q_{2b} \left | \exp \left \{ -( \frac{ i}{\hbar})\left [ \frac{ 1}{2m}\hat{p}_2^2 +U_2(q_2)
-EV_2(q_2)
\right ] \tau \right \} \right | q_{2a} \right > \nonumber \\
\label{200}
\end{eqnarray}
The RHS of Eq. \ref{200} is written in terms of path integral
\cite{fu}.
\begin{eqnarray}
\left <q_{1b},\, q_{2b} \left |\exp \left \{-i  \frac{ \hat{H}_T\tau}{\hbar}\right \}
\right |q_{1a}, \, q_{2a} \right >\mbox{\hspace{3in}} \nonumber \\= 
 \int {\cal D }q_1{\cal D}p_1 \exp \left \{ ( \frac{ i}{\hbar}\int _{0}^{\tau}
\left [ p_1\dot{q_1} -\left ( \frac{ \hat{p}_1^2}{2m}+U_1(q_1) -EV_1(q_1) \right )
\right ] d\tau \right \}\nonumber \\
\times\int {\cal D }q_2{\cal D}p_2\exp \left \{ ( \frac{ i}{\hbar}\int _{0}^{\tau}
\left [ p_2\dot{q_2} -\left ( \frac{ \hat{p}_2^2}{2m}+U_2(q_2) -EV_2(q_2) \right )
\right ] d\tau \right \}
\label{201}
\end{eqnarray}
The parameter $\tau $ is arbitrary and one can obtain physically meaningful
quantity out of Eq. \ref{200} by integrating  over $\tau $ from $0$ to
$\infty$ .
\begin{eqnarray}
\left <q_{1b},\, q_{2b} \left |   \frac{\hbar}{ \hat{H}_T}
\right | q_{1a}, \, q_{2a} \right >_{\mbox{Semi-classical}}\mbox{\hspace{3.6in}}\nonumber \\
=\int _0^\infty d\tau
\left < q_{1b} \left | \exp \left \{ -( \frac{ i}{\hbar})\left [ \frac{ 1}{2m}\hat{p}_1^2 +U_1(q_1) -EV_1(q_1)
\right ] \tau \right \} \right | q_{1a} \right > \nonumber \\
\times   \left < q_{2b} \left | \exp \left \{ -( \frac{ i}{\hbar})\left [ \frac{ 1}{2m}\hat{p}_2^2 +U_2(q_2)
-EV_2(q_2)
\right ] \tau \right \} \right | q_{2a} \right >\nonumber \\
 \label{2002}
\end{eqnarray}
The meaning of RHS of Eq. \ref{2002} has been explained in \cite{fu}.
And the LHS of it is written as, 
\begin{equation}
\left <q_{1b},\, q_{2b} \left |   \frac{\hbar}{ \hat{H}_T}
\right | q_{1a}, \, q_{2a} \right > =
\left <q_{1b},\, q_{2b} \left |   \frac{\hbar}{ \hat{H}-E}
\right | q_{1a}, \, q_{2a} \right > \frac{1}{V_1(q_{1a}) +V_2(q_{2a})}
\label{202}
\end{equation}
with the completeness relation
\begin{equation}
\int\left | q_1\  q_2 \right >\frac{dV}{V_1(q_1)+V_2(q_2)}\left < q_1 \ q_2 \right | =1
\label{203}
\end{equation}

Now we   show that the Hamiltonian  of a particle in a non-central potential is reduced to a 
separable Hamiltonian of the above kind. Let us start with the non-central system  written in Eq. \ref{pot}
\begin{equation}
H= \frac{P_r^2}{2m}+\frac{P_\theta^2}{2mr^2} +\frac{P_\phi^2}{2mr^2 sin^2\theta}-\frac{ \alpha }{r} + \frac{ \beta }{r^2\sin^2 \theta }
+ \frac{ i\gamma \cos\theta }{r^2\sin^2 \theta}
\label{23}
\end{equation}
where 
 $ \beta  = \frac{B\hbar^2}{2m}; \ \ \gamma  = \frac{C\hbar^2}{2m}.$
 This system will be reduced to separable system of Liouville type in parabolic coordinate 
system.
To express the Hamiltonian in parabolic coordinates $(\xi, \eta, \phi)$,
it is useful to first  express this potential $ V(r,\theta)$ in cylindrical coordinates
$(\rho, \phi, z)$. 
 In cylindrical coordinate the potential looks like,
\begin{equation}
V(\rho,z) = -\frac{\alpha }{\sqrt{\rho^2 +z^2}} + \frac{\beta }{\rho^2} +\frac{i \gamma z}{\rho^2
\sqrt{\rho^2+z^2}};\ \ \ \ \ \rho^2= x^2 + y^2
\label{25}
\end{equation}  
The parabolic coordinates are expressed in terms of cylindrical coordinates as
\begin{eqnarray}
\xi &=& \frac{1}{2}\left ( \sqrt{\rho^2+z^2} -z \right ) ; \nonumber\\
\eta &=&\frac{1}{2}\left (\sqrt{\rho^2+z^2} + z \right ); \ \ \ \  
\phi = \phi \ . \label{26}
\end{eqnarray}
Now the potential in Eq. \ref{25} in terms of these parabolic coordinates,
is
\begin{equation}
V(\xi,\eta) = -\frac{\alpha }{\xi+\eta}+ \frac{ \beta }{4\xi\eta} + \frac{
i \gamma (\eta-\xi)}{4\eta\xi(\eta+\xi)}\label{27}
\end{equation}
and the Hamiltonian in parabolic coordinate is 
 written as
\begin{equation}
H(\xi,\eta,\phi) = \frac{ 1}{2m(\xi+\eta)}\left [\xi \hat{p}_{\xi}^2 +
\eta \hat{p}_\eta^2 \right ] +\frac{ 1}{8m\eta\xi} \hat{p}_\phi^2 +V(\xi,\eta)
\label{28}
\end{equation}
We define variables $u, v$
\begin{eqnarray}
\xi &=& \frac{ 1}{4} u^2 ;\ \ \ \ \ 0\leq u<\infty \nonumber\\
\eta &=&\frac{ 1}{4} v^2 ;\ \ \ \ \ 0\leq v<\infty \label{29}
\end{eqnarray}
and perform a canonical transformation 

\begin{equation}
\sqrt{\xi}\hat{p}_\xi =\hat{ p}_u \ ; \
\sqrt{\eta}\hat{p}_\eta = \hat{p}_v \label{210}
\end{equation}

to simplify
the kinetic term in $H$ in Eq. \ref{28} as
\begin{equation}
H(u,v,\phi) = \frac{ 4}{u+v}\left \{ \frac{ 1}{2m} \left [\hat{p}_u^2 +\hat{p}_v^2
+(\frac{ 1}{ u^2}+ \frac{ 1}{v^2})\hat{p}_\phi^2 \right ] -\alpha  + \frac{ \beta +i \gamma }{u^2}
+ \frac{ \beta -i \gamma }{v^2} \right \} \label{211}
\end{equation}
This is further written compactly as 
\begin{equation}
H(u,v,\phi) = \frac{ 4}{u^2+v^2} \left \{ \frac{ 1}{2m} \left [\hat{ p}_u^2
+\hat{p}_v^2 +\frac{ 1}{u^2}\hat{p}_{\phi_1}^2 +\frac{
1}{v^2}\hat{p}_{\phi_2}^2 \right ] -\alpha  \right \}
\label{212}
\end{equation}
where
\begin{equation}
\hat{p}_{\phi_1}^2 = \hat{p}_{\phi}^2 + 2m(\beta +i \gamma );\;\;\;\; 
\hat{p}_{\phi_2}^2 = \hat{p}_{\phi}^2 + 2m(\beta -i \gamma ) 
\label{213}
\end{equation}
Note $\hat{p}_{\phi_1}$ and $ \hat{p}_{\phi_2}$ are not Hermitian but complex conjugate to each other.
This Hamiltonian is still not a separable one of Liouville type.
We  further consider the total Hamiltonian $H_T (=H-E)$ with $E=-2m\omega^2$ for the bound state case ($E< 0$),
\begin{eqnarray}
\hat{H}_T &=& \frac{ 1}{2m} \left [\hat{p}_u^2 
+\hat{p}_v^2 +\frac{ 1}{u^2}\hat{p}_{\phi_1}^2 +\frac{
1}{v^2}\hat{p}_{\phi_2}^2 \right ] -\alpha + \frac{1}{2}m\omega^2 (u^2+v^2) 
\label{214}
\end{eqnarray}
Now we introduce the components of 2-dimensional vectors $\vec{u} $ and $\vec{v}$ as
\begin{eqnarray}
\vec{u} &=& ( u_1, u_2) = \left ( u\cos\phi_1 , u\sin\phi_1 \right )
\nonumber \\
\vec{v} &=& ( v_1, v_2) = \left ( v\cos\phi_2 , v\sin\phi_2 \right )
\label{215}
\end{eqnarray}

to have
\begin{equation}
\vec{p_u}^2 = \hat{p}_u^2 + \frac{ 1}{u^2}\hat{p}_{\phi_1}^2;
\;\;\;\;\;\;\;
\vec{p_v}^2 = \hat{p}_v^2 + \frac{ 1}{v^2}\hat{p}_{\phi_2}^2;\label{216}
\end{equation}
Putting all these in 
Eq. \ref{214} we obtain 
\begin{equation}
\hat{H}_T = \frac{ 1}{2m} \vec{p}_u^2 +\frac{ m\omega^2}{2}\vec{u}^2 +
\frac{ 1}{2m} \vec{p}_v^2 + \frac{ m\omega^2}{2}\vec{v}^2 -\alpha 
\label{217}
\end{equation}
This is the Hamiltonian which is separable of Liouville type. 
Thus, the Hamiltonian for the non-central
potential has been reduced to that of two 2-dimensional oscillators (apart from some
constant shift in ground state energy ).

Now
by using Eq. \ref{200} for this separable Hamiltonian, we  find the path integral 
for the non-Hermitian non-central potential exactly as
\begin{eqnarray}
\left < \vec{u}_b ,\, \vec{v}_b \left |\exp \left [-i \frac{
\hat{H}_T\tau}{\hbar} \right ] \right |\vec{u}_a ,\, \vec{v}_a \right > =
e^{\frac{ i\alpha\tau}{\hbar}}
 \left < \vec{u}_b \left |\exp \left [ - (\frac{
i}{\hbar})\left ( \frac{ \vec{p}_u^2}{2m} + \frac{ m\omega^2}{2}\vec{u}^2 \right )\tau \right ] \right
| \vec{u}_a \right >\nonumber \\
\times \left < \vec{v}_b \left |\exp \left [ - (\frac{
i}{\hbar})\left ( \frac{ \vec{p}_v^2}{2m} + \frac{ m\omega^2}{2}\vec{v}^2 \right )\tau \right ] \right
| \vec{u}_a \right >\nonumber \\
=e^{(\frac{ i\alpha\tau}{\hbar})}{ \left (\frac{i m\omega}{2\pi i\hbar\sin\omega\tau}\right )}^{
\frac{ 4}{2}}\exp \left \{ \frac{ im\omega}{2\hbar\sin\omega\tau} \left [
\left ( \vec{u}_b^2 + \vec{v}_b^2 +\vec{u}_a^2 +\vec{v}_a^2 \right
)\cos\omega\tau - 2\vec{u}_b\cdot\vec{u}_a -2\vec{v}_b\cdot\vec{v}_a \right ] \right
\}\nonumber \\
\label{pt}
\end{eqnarray}
where  the exact result  for one dimensional simple harmonic oscillator has been used \cite{fey}.
\begin{equation}
\left < q_b \left | \exp \left [ - \frac{ i}{\hbar} \left ( \frac{
\hat{p}^2}{2m}
+ \frac{ m\omega^2 q^2}{2} \right  )\tau \right ] \right | q_a \right > =
\left ( \frac{ m\omega}{2\pi i \hbar\sin \omega\tau} \right )^{ \frac{ 1}{2}}
\exp \left \{ \frac{ im\omega}{2\hbar\sin \omega\tau} \left [ (q_a^2 +q_b^2)
\cos \omega\tau - 2q_aq_b \right ] \right \}
\end{equation}

The Eq. \ref{pt} 
 contains the arbitrary parameter
$\tau$ and has to be eliminated to obtain physically meaningful quantity.
This can be done by integrating over $\tau$ from $0$ to $\infty$ in both side
of the Eq. \ref{pt}. When we integrate over $\tau$ in the LHS of the
Eq. \ref{pt},
it is nothing but  the Green's functions of the operator $\frac{1}{\hat{H} -E}$ as discussed 
at beginning of this section.
And the integration in the RHS can be done in a  straightforward manner \cite{fu}.
Thus we obtain the explicit Green's functions  for the system of non central non-Hermitian
potential.

\section{Reality of the Spectrum}
\label{s3}

Since  this system with non-Hermitian, non-central potential is equivalent to two 2d SHOs, conjugate to each other
we define  the creation and annihilation operators for this theory as

 \begin{eqnarray}
a_k = \frac{ 1}{\sqrt{2}} \left [ \sqrt{ \frac{ m\omega}{\hbar}}u_k + \frac{ i}{
\sqrt{m\omega \hbar}}\hat{p}_{u_k} \right ] \nonumber \\ 
\tilde{a}_k = \frac{ 1}{\sqrt{2}} \left [ \sqrt{ \frac{ m\omega}{\hbar}}v_k + \frac{ i}{
\sqrt{m\omega \hbar}}\hat{p}_{v_k} \right ]  
\label{31}
\end{eqnarray}
where $k= 1, 2$  to obtain the energy levels
for this system of non-central potential in a simple algebraic way.
The total Hamiltonian in Eq. \ref{217} is written in terms of these creation
and annihilation  operators as follows:
\begin{equation}
\hat{H}_T = \hbar\omega \left [ \sum _{k=1} ^ 2 \left (a_k^{\dagger}a_k +
\tilde{a}_k^{\dagger}\tilde{a}_k \right ) +2 \right ] -\alpha
\label{32}
\end{equation}
and the conjugate momentum variables are written as
\begin{eqnarray}
\hat{p}_{\phi_1} = i\hbar \left [ a_1^\dagger a_2 - a_2^\dagger a_1 \right
]\nonumber \\
\hat{p}_{\phi_2} = i\hbar \left [ \tilde{a}_1^\dagger \tilde{a}_2 -
\tilde{a}_2^\dagger\tilde{
a}_1
\right ]\label{33}
\end{eqnarray}
We further perform an unitary transformation of the following type,
\begin{eqnarray}
a_1 = \frac{ 1}{\sqrt{2}} \left ( b_1 - ib_2 \right )\nonumber \\
a_2 = \frac{ 1}{\sqrt{2}} \left ( -ib_1 + b_2 \right )
\label{34}
\end{eqnarray}
and similar transformations for $\tilde{a}_1 , \tilde{a}_2$ also
in Eqs. \ref{32} and \ref{33} to get,
\begin{equation}
\hat{H}_T = \hbar\omega \left [\sum_{k=1}^2 \left (b_k^\dagger b_k +
\tilde{b}_k^\dagger \tilde{b}_k \right )+2 \right ] -\alpha \
\label{35}
\end{equation}
and 
\begin{eqnarray}
\hat{p}_{\phi_1} = \hbar \left [b_1^\dagger b_1 - b_2^\dagger b_2 \right ]
\nonumber \\
\hat{p}_{\phi_2} = \hbar \left [\tilde{b}_1^\dagger\tilde{ b}_1 - \tilde{b}_2^\dagger\tilde{ b}_2 \right ]
\label{36}
\end{eqnarray}
 The number operators,
$
n_k = b_k^\dagger b_k \ , \
\tilde{n}_k = \tilde{b}_k^\dagger \tilde{b}_k $ are defined for this system.
In terms of number operators the total Hamiltonian in Eq. \ref{35} is now written as
\begin{eqnarray}
\hat{H}_T &=& \hbar\omega \left [ n_1 + n_2 + \tilde{n}_1 +\tilde{n}_2 +2
\right ] -\alpha \nonumber \\
&=& 2\hbar\omega \left [n_2 +\tilde{n}_2+1+ \frac{ \hat{p}_{\phi_1} +\hat{p}_{\phi_2}}{\hbar}
\right ] -\alpha
\end{eqnarray}
Now considering
$\left [\hat{p}_{\phi_1} + \hat{p}_{\phi_2} \right ] \phi_{phy}\equiv
\lambda \phi_{phy}$, the physical state condition is \cite{fu},
\begin{equation}
\left [ 2(n_2+\tilde{n}_2+1) \hbar\omega + \omega \lambda  
-\alpha \right ] \phi_{phy} = 0 \label{kk1}
\end{equation} 
Hence the energy level can be written in terms of $\lambda $ as,
\begin{equation}
E_{n_2,\tilde{n}_2,\lambda } \equiv -2m\omega^2 = -\frac{ 2m\alpha^2}{\left [2(n_2+\tilde{n}_2+1)\hbar +\lambda \right ]^2}.
\label{kk3}
\end{equation}
  $\lambda
$ can be calculated easily using $\left [\hat{p}_{\phi_1} + \hat{p}_{\phi_2} \right ] \phi_{phy}\equiv
\lambda \phi_{phy}$ and Eq. \ref{213}
as 
\begin{eqnarray} 
\lambda  &=& \hbar \left [ \left (\nu^2 +B+iC \right )^{ \frac{ 1}{2}} +
\left (\nu^2+B-iC \right )^{ \frac{ 1}{2}} \right ] \nonumber\\ 
&=&  \left [ \left (\nu^2\hbar^2 +2m(\beta+i\gamma) \right )^{ \frac{ 1}{2}} +
\left (\nu^2\hbar^2+2m(\beta-i\gamma) \right )^{ \frac{ 1}{2}} \right ]
\end{eqnarray}
where $\nu$ is non-negative integer and $\lambda $ is real as $\lambda=\lambda^* $.
Therefore, the complete real bound state spectrum for the problem is
\begin{eqnarray}
E_{n_2,\tilde{n}_2,\nu}
&=&  \frac{- m \alpha^2}{2\hbar^2 \left [
n_2+\tilde{n}_2 +1 +\frac{\sqrt{\nu^2 +B+iC} + \sqrt{\nu^2 +B-iC}}{2} \right ]^2}\nonumber \\
 &= & \frac{- m \alpha^2}{2 \left [
(n_2+\tilde{n}_2 +1)\hbar +\frac{\sqrt{\nu^2\hbar^2 +2m(\beta+i\gamma)} + \sqrt{\nu^2\hbar^2 +2m(\beta-i\gamma)}}{2} \right ]^2}
\label{ene}
\end{eqnarray}

The corresponding  result for the real potential  agrees with that of in Refs. \cite{ast1,che}  where energy spectrum has 
been calculated by solving Schroedinger equation using complicated KS transformation \cite{duru,ks}.

\section{ Conclusion}
\label{s4}
 The  Hamiltonian corresponding to the PT symmetric non-Hermitian non-central  potential
in Eq. \ref{pot} has been mapped into a Hamiltonian of two 2d harmonic
oscillators by choosing appropriate coordinate system and using a suitable canonical 
transformation. Next we have calculated the Green's functions for the
system using path integral method for this separable Hamiltonian of
Liouville type. The exact spectrum are calculated by 
 writing this Hamiltonian  in terms of
creation and annihilation operators of 2d SHO. The entire spectrum is real for any real values of the 
parameters $\alpha,\ \beta $ and $\gamma $ indicating that system is always in unbroken phase.


\vspace{.2in}

{\bf Acknowledgments}:
BPM acknowledge the financial support from the Department of Science and Technology (DST), Govt. of India under SERC project sanction grant No. SR/S2/HEP-0009/2012. 

\input{referenc}

\end{document}

%% file: referenc.tex


%% file: path_nh1_1_.bbl
\begin{thebibliography}{99.}
\bibitem{fey} R.P. Feynman and A.R. Hibbs, {\it Quantum Mechanics and
Path Integral } (McGraw Hill, New York, 1965)
\bibitem{adas} See for example, A Das, {\it Field Theory: A Path Integral Approach} ( World Scientific)
\bibitem{pt} C. M. Bender and S. Boettcher {\em Phys.Rev.Lett.} {\bf 80}, 5243 (1998);
 A. Mostafazadeh, {\em Int. J. Geom. Meth. Mod. Phys.} {\bf 7}, 1191(2010) and references therein;
 C. M. Bender, {\em Rep.Prog. Phys.} {\bf 70}, 947 (2007) and references therein.

\bibitem{ptp} G. Levai, J. Phys. A 41 (2008) 244015;
 C. M. Bender, G. V. Dunne, P. N. Meisinger, M. Simsek Phys. Lett.A 281 (2001)311-316.
 C. M. Bender, David J. Weir J. Phys. A 45 (2012)425303;
 B. P. Mandal, B. K. Mourya, and R. K. Yadav (BHU),Phys. Lett. A 377, 1043 (2013);
 B. P. Mandal, B. K. Mourya, K, Ali and A. Ghatak, arXiv. 1509.07500 ( Press, Ann. of Phys. (2015));
 M. Znojil J. Phys.A 36(2003) 7825;
 C. M. Bender, G. V. Dunne, P. N. Meisinger, M. Simsek Phys. Lett. A 281 (2001)311-316.
 

\bibitem{pto}  C. E. Ruter, K. G. Makris, R. El-Ganainy, D. N. Christodulides, M. Segev, and D. Kip, Nature (London) Phys. 6, 192 (2010);
 A. Guo, G. J. Salamo, Phys. Rev. Lett. 103, 093902 (2009);
 C. M. Bender, S. Boettcher and P. N. Meisinger, J. Math. Phys. 40 2201 (1999);
 C. T. West, T. Kottos, and T. Prosen,Phys. Rev. Lett. 104, 054102(2010);
 A. Nanayakkara Phys. Lett. A 304, 67 (2002)

\bibitem{pta}
 Z. H. Musslimani, K. G. Makris, R. El-Ganainy, and D. N. Christodou
lides, Phys. Rev. Lett. 100, 030402 (2008);M. V. Berry, Czech. J. Phys. 54, 1039 (2004);
 W. D. Heiss,Phys. Rep. 242, 443 (1994).
 
\bibitem{pta0} C. M. Bender, D. C. Brody, H. F. Jones, {\it Phys.Rev. D} {\bf 70}, 025001 (2004).
 C. M. Bender, D. C. Brody, J. Caldeira, B. K. Meister, {\em arXiv} {\bf 1011.1871} (2010).
 S. Longhi {\em Phys. Rev. B}  {\bf 80}, 165125 (2009).
 B. F. Samsonov {\em J. Phys. A} {\bf 43}, 402006 (2010); {\em Math. J. Phys. A}: {\em Math. Gen.} {\bf 38}, L571 (2005).

\bibitem{pta1} B. Basu-Mallick and B.P. Mandal, Phys. Lett. A 284, 231 (2001);
 B. Basu-Mallick, T. Bhattacharyya and B. P. Mandal Mod. Phys. Lett. A 20, 543 (2004);
 
\bibitem{pta2} A. Ghatak, R. D. Ray Mandal, B. P. Mandal,Ann. of Phys.336, 540 (2013);
 A. Ghatak, J. A. Nathan, B. P. Mandal, and Z. Ahmed,J. Phys. A: Math. Theor.45, 465305 (2012);
A. Mostafazadeh, Phys. Rev. A 87, 012103 (2013);
 L. Deak, T. Fulop Ann. of Phys. 327, 1050 (2012);
 J. N. Joglekar, C. Thompson, D. D. Scott and G Vemuri, Eur. Phys. J. Appl. Phys. 63, 30001 (2013) ;
 M. Hasan, A. Ghatak, B. P. Mandal,Ann. of Phys. 344, 17 (2014);
 A. Ghatak and B. P. Mandal J. Phys. A: Math. Theor. 45, 355301 (2012); 
 S. Longhi, {\em J. Phys. A: Math. Theor.} {\bf 44}, 485302 (2011).

 
\bibitem{pta3} B. P. Mandal and S S. Mahajan , Comm. Theor. Phys. 64, 425(2015) 
; S Dey and A Fring Phys. Rev A 88, 022116 (2013);  C. M. Bender, D. W. Hook, P. N. Meisinger, Q. Wang, {\it Phys.Rev.Lett.} {\bf 104}, 061601 (2010); C. M. Bender, D. W. Hook, K. S. Kooner  {\it J.Phys.A} {\bf 43},  165201, (2010).
\bibitem{pta4} A. Khare and B. P. Mandal, {\em Phys.Lett.} {\bf A272}, 53 (2000);
  B. P. Mandal, S. Gupta, {\em Mod.Phys.Lett. A} {\bf 25}, 1723 (2010);
 B. P. Mandal, {\em Mod. Phys. Lett. A} {\bf 20}, 655(2005);
 B. P. Mandal and A. Ghatak, J. Phys. A: Math. Theor. 45, 444022 (2012);
M Znojil, Ann. of Phys. 361, 226 (2015).

 
 \bibitem{fu} K. Fujikawa, Nucl. Phys. {\bf B 484}, 495, (1997);
  \bibitem{bpm} B P Mandal, Int. J of Mod. Phys A 15, 1225 (2000).

  
  \bibitem{hat} H. Hartmann, Theor. Chim. Acta {\bf 24} 201 (1972).


\bibitem{ast1} M. Kibler and C. Campigotto, Int. J. of Quantum Chemistry, {\bf 45}, 209(1993). 


\bibitem{che} L. Chetouani, L. Guechi and T. F. Hammann, J. Math. Phys.
{\bf 33} 3410 (1992).

\bibitem{duru} L. H. Duru and H. Kleinert,  Phys. Lett. {\bf B84}, 185(1979).
\bibitem{ks} P. Kustaanheimo and E. Stiefel, J. Reine Angew Math. {\bf 218}
204(1965).

\end{thebibliography}
